\documentclass[aps,prb,twocolumn]{revtex4-1}

\usepackage[utf8]{inputenc}
\usepackage{amsmath}
\usepackage{graphicx}
\usepackage{bbold}
\usepackage{soul}
\usepackage{xcolor}
\usepackage{tabularx}
\usepackage{multirow}
\usepackage{makecell}
\usepackage[normalem]{ulem}

\begin{document}

\title{Swapping exchange and spin-orbit induced correlated phases in proximitized Bernal bilayer graphene}

\author{\surname{Yaroslav} Zhumagulov$^{1}$}
\email[Emails to: ]{iaroslav.zhumagulov@ur.de}
\author{\surname{Denis} Kochan$^{2,1}$}
\email[Emails to: ]{denis.kochan@savba.sk}
\author{\surname{Jaroslav} Fabian$^{1}$}
\email[Emails to: ]{jaroslav.fabian@ur.de}

\affiliation{%
  $^1$Institute for Theoretical Physics, University of Regensburg,
 93040 Regensburg, Germany\\
 $^2$Institute of Physics, Slovak Academy of Sciences, 84511 Bratislava, Slovakia
 }%

\date{\today}

\begin{abstract}
Ex-so-tic van der Waals heterostructures take advantage
of the electrically tunable layer polarization to swap proximity exchange and spin-orbit coupling in the electronically active region. Perhaps the simplest example is Bernal bilayer graphene (BBG) encapsulated by a layered magnet from one side and a strong spin-orbit material
from the other. Taking WS$_2$/BBG/Cr$_2$Ge$_2$Te$_6$
as a representative ex-so-tronic device, we employ
realistic \emph{ab initio}-inspired Hamiltonians and
effective electron-electron interactions to investigate 
the emergence of correlated phases within the random phase approximation. 
We find that exchange and spin-orbit coupling
induced Stoner and intervalley coherence instabilities can be swapped for a given doping level, allowing to 
explore the full spectrum of correlated phases within a single device.  
\end{abstract}

\maketitle

\section{Introduction.} 

While a variety of novel electronic correlation effects have initially been discovered
in twisted graphene structures~\cite{Dean2013, Kim2016, Cao2018_1, Cao2018_2}, 
recent demonstrations of quarter and half-metallic states~\cite{Zhou2021_1, PhysRevB.105.L081407, PhysRevB.106.155115, PhysRevB.107.L121405, arxiv.2305.04950, Geisenhof2021} and superconductivity~\cite{Zhou2021_2, PhysRevLett.127.187001, PhysRevLett.127.247001, PhysRevB.105.134524, Chatterjee2022, PhysRevB.106.155115, PhysRevB.105.075432, arxiv.2203.09083} in rhombohedral trilayer graphene, as well as the observation of isospin magnetism and spin-polarized superconductivity in Bernal bilayer graphene (BBG)~\cite{Zhou2022, Zhang2023, Seiler2022, PhysRevB.105.L100503, delaBarrera2022, arxiv.2303.00742, PhysRevB.105.L201107, PhysRevB.106.L180502, PhysRevB.107.L201119}, are clear evidence for correlated physics
without moiré patterns. The common feature, boosting the electron-electron interactions in these
systems is the presence of van Hove singularities (vHS) near the charge neutrality point~\cite{Li2009, Bistritzer2011, PhysRevB.82.035409}. Particularly attractive is the tunability of the correlated phases by a displacement field, which can shift the
electronic levels of the vHS.~\cite{Zhou2021_1, Zhou2021_2, Zhou2022, Zhang2023}.

\begin{figure}[!h]
    \centering
    \includegraphics[width=.5\textwidth]{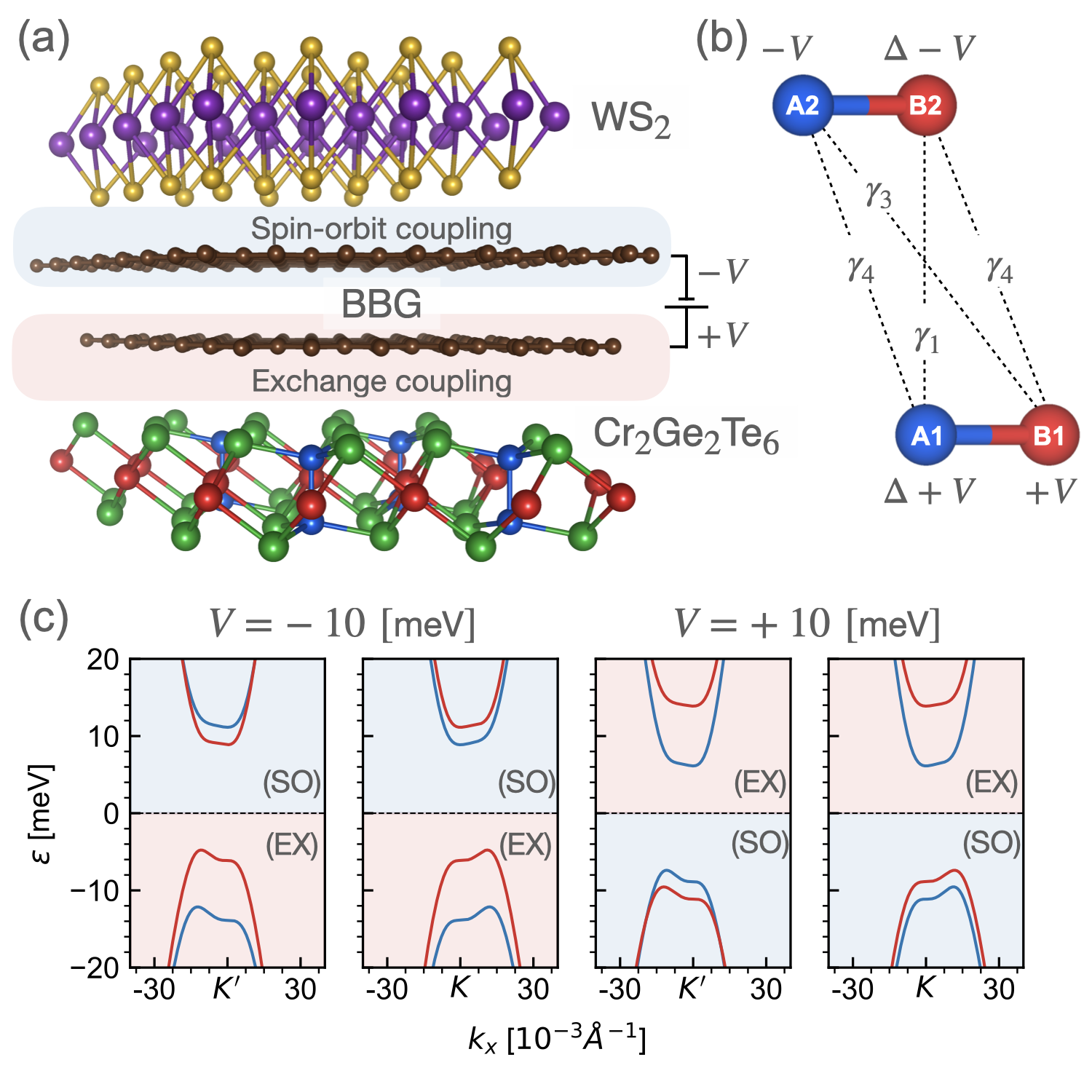}
    \caption{
    (a) Scheme of an ex-so-tic heterostructure comprising BBG encapsulated by WS$_2$ and CGT which  proximitize the BBG by SO (WS$_2$) and EX (CGT) interactions. 
    (b) BBG unit cell with relevant interlayer orbital hoppings and on-site energies. The intralayer nearest-neighbor hopping $\gamma_0$ is not indicated. 
    The colors distinguish $A$ and $B$ sublattices within the two layers. 
    (c) Calculated single-particle low-energy electronic dispersions of WS$_2$/BBG/CGT near $K$ and $K^\prime$ points obtained from $h({\bf k},\tau)$ for $V=\pm 10$\,meV;
    red bands are spin-polarized up, blue are spin-polarized down; the spin quantization axis $z$ is perpendicular to the layers.
    }  
    \label{fig:scheme}
\end{figure}

The electronic states of 2D materials in van der Waals heterostructures can be also affected by proximity effects~\cite{Sierra2021}. Relatively strong spin interactions, for example, can be induced in materials
such as graphene which exhibits weak spin-orbit coupling~\cite{PhysRevB.80.235431, PhysRevB.85.115423, PhysRevB.82.245412}. Indeed, proximity-induced spin-orbit (SO) and 
exchange (EX) interactions have been predicted theoretically~\cite{Gmitra2013, PhysRevB.99.085411, PhysRevB.104.195156, Gmitra2015, Gmitra2016, PhysRevB.104.075126, PhysRevLett.125.196402, PhysRevB.105.115126, PhysRevB.94.155441, PhysRevB.77.115406, Kochan2017, PhysRevB.104.075126, PhysRevLett.125.196402, PhysRevB.105.115126, ZollnerTwist} and confirmed experimentally~\cite{Garcia2018, Island2019, Hoque2021, Ghiasi2017, Ghiasi2019, Safeer2019, Herling2020, Wakamura2019, Wakamura2020, InglaAyns2021, Kaverzin2022, Karpiak2019, Rybkin2018, PhysRevB.91.235431, PhysRevX.2.041017} in BBG-based heterostructures. 
Specifically, valley-Zeeman, Kane-Mele, and Rashba SO couplings have been
shown to emerge~\cite{Gmitra2013, Garcia2018, Island2019, Hoque2021, Ghiasi2017, Ghiasi2019, Safeer2019, Herling2020, Wakamura2019, Wakamura2020, InglaAyns2021, Kaverzin2022, PhysRevB.99.085411,PhysRevB.104.195156, Gmitra2015, Gmitra2016, PhysRevB.104.075126, PhysRevLett.125.196402, Gmitra2013, PhysRevB.105.115126}, along with (anti)ferromagnetic EX couplings~\cite{ZollnerTwist, PhysRevB.94.155441, PhysRevB.77.115406, PhysRevB.99.195452, PhysRevB.104.075126, PhysRevLett.125.196402, PhysRevB.105.115126, Kaverzin2022,Karpiak2019, ZollnerTwist, PhysRevB.91.235431, PhysRevX.2.041017}, typically causing spin 
splittings on the meV scale. 

Perhaps the most striking manifestation of the proximity effects is the possibility to swap the spin couplings, EX and SO, by a displacement field, whereby changing the spin Hamiltonian in the active layer. This effect arises due to the interplay of the short-range proximity interactions and layer polarization by the field. An experimentally 
relevant materials system to consider is that of BBG encapsulated from one side by a strong spin-orbit material, such as WS$_2$, and a magnetic semiconductor, such as Cr$_2$Ge$_2$Te$_6$ (CGT), from the other side. The resulting heterostructure, 
termed ex-so-tic~\cite{PhysRevLett.125.196402}, allows to imprint selectively either 
SO or EX coupling onto the Bloch states of the BBG.  

Given that BBG can host correlated phases, is it possible to swap
between the phases induced by EX and SO coupling? As we show here, ex-so-tronic
devices can indeed supply on-demand correlated phases, allowing to swap the 
effective single-particle excitation Hamiltonians enabled by the layer polarization 
effect. Although our conclusions are more general, we specifically study 
WS$_2$/BBG/CGT, for which an effective orbital and spin-interaction Hamiltonian has
been derived from DFT~\cite{PhysRevB.104.075126, PhysRevLett.125.196402}. 
Employing random-phase approximation (RPA)~\cite{PhysRev.82.625, PhysRev.85.338, PhysRev.92.609, PhysRevLett.101.087004, Graser2009, PhysRevB.83.100515}, we first examine the particle-hole instabilities of pristine BBG, whose correlated phase diagram features  intervalley coherence (IVC) ~\cite{PhysRevLett.103.216801} and Stoner instabilities~\cite{PhysRevB.105.134524, Chatterjee2022}. 
The spin interactions induced by proximity SO (see the latest works  ~\cite{PhysRevB.107.L201119,wang2023electrical,koh2023correlated,zhumagulov2023emergent}) and EX couplings remove the spin and valley degeneracies of
IVC and Stoner phases and cause a plethora of emergent spin-valley correlated phases. In 
ex-so-tic heterostructures, such as in here employed WS$_2$/BBG/CGT, the phases can be effectively 
swapped, while the evidence of the interplay of SO and EX interactions can be seen 
in the single-particle (Hartee-Fock) excitation spectra which we also calculate. 


\section{Model.} 

\begin{table*}[t]
\centering
\begin{tabular}{||c|c|c|c|c|c|c|c|c||}
\hline 
$\gamma_0$[eV] &
$\gamma_1$[eV] &
$\gamma_3$[eV]&
$\gamma_4$[eV]&
$\Delta$[meV]&
$\lambda^{A_2}_I$[meV]&
$\lambda^{B_2}_I$[meV] &
$\lambda^{A_1}_X$[meV]&
$\lambda^{B_1}_X$[meV] \\
\hline
2.432 
& 0.365 
 & -0.273
 & -0.164
 & 8.854
 & 1.132 
& -1.132 
 & -3.874
 & 3.874 \\
\hline
\end{tabular}
\caption{
The numerical values for the parameters of the low-energy single-particle Hamiltonian of WS$_2$/BBG/CGT heterostructure are taken from the \emph{ab initio} results of Ref.~\cite{PhysRevLett.125.196402}.}
\label{tab:bbg-prox}
\end{table*}

The orbital degrees of freedom of BBG are modeled by a realistic Bloch Hamiltonian~\cite{PhysRevB.85.115423, PhysRevLett.125.196402}:
\begin{equation}
\centering
    \hat{h}_0(\mathbf{k},\tau)\
=\begin{pmatrix}
  \Delta+V & \gamma_0 f(\mathbf{k})&\gamma_4 f^{*}(\mathbf{k})&\gamma_1 \\ 
  \gamma_0 f^{*}(\mathbf{k}) & +V&\gamma_3 f(\mathbf{k})&\gamma_4 f^{*}(\mathbf{k})\\
  \gamma_4 f(\mathbf{k})&\gamma_3 f^{*}(\mathbf{k})&-V&\gamma_0 f(\mathbf{k})\\
  \gamma_1&\gamma_4 f(\mathbf{k})&\gamma_0 f^{*}(\mathbf{k})&\Delta-V\\
\end{pmatrix}
\label{eq:kin-ham}
\end{equation}
in the basis of $p_z$ orbitals ordered as ($A_1, B_1, A_2, B_2$). Parameters $\gamma$
denote intra- ($\gamma_0$) and interlayer ($\gamma_1$, $\gamma_3$, $\gamma_4$) hoppings, see
Figs.~\ref{fig:scheme}~(b)~and~(c). 
The displacement field is accounted for by on-site energy $V$, while the asymmetry between A and B sublattice
is quantified by $\Delta$. Momenta $\mathbf{k}=(k_x, k_y)$ are measured from $K$ and $K'$ valleys, while, 
$f(\mathbf{k})=-\left(\sqrt{3}a/2\right)\left(\tau_x-ik_y\right)$ is the linearized nearest-neighbor tight-binding function; $a=2.46\,\AA$ is graphene's lattice constant and $\tau=\tau_{K/K'}=\pm1$ is the valley index. Since the SO coupling of pristine BBG is weak---about $24~\mu$eV~\cite{PhysRevB.85.115423}---we do not include it in the Hamiltonian. 

To model proximitized BBG in a WS$_2$/BBG/CGT ex-so-tic heterostructure, 
we need to include SO (due to WS$_2$) and EX (due to CGT) couplings, which were shown 
by \emph{ab initio} calculations and weak-antilocalization measurements~\cite{PhysRevB.105.115425} to be sizeable, on the order of 1~meV~\cite{PhysRevB.104.075126, PhysRevLett.125.196402}. The corresponding proximity induced spin interactions are described
near $K$ and $K'$ valleys by the Hamiltonian $\hat{h}_{\text{prox}}(\tau)=\sum_{l} \hat{h}^{l}_{\text{vz}}(\tau)+\hat{h}^{l}_{\text{ex}}(\tau)$, for valley-Zeeman (vz) spin-orbit and exchange (ex) coupling ~\cite{PhysRevB.105.115126, Gmitra2016,Kochan2017,Gmitra2013}:
\begin{align}
    \hat{h}^{l}_{\text{vz}}(\tau)+\hat{h}^{l}_{\text{ex}}(\tau)&= 
    \begin{pmatrix}
        (\tau\,\lambda_{\text{vz}}^{A_l}-\lambda_{\text{ex}}^{A_l})\,s_z&0\\
        0&-(\tau\,\lambda_{\text{vz}}^{B_l}+\lambda_{\text{ex}}^{B_l})\,s_z
    \end{pmatrix},
    \label{eq:prox}
\end{align}
parameterized by the corresponding sublattice and layer-resolved couplings $\lambda^{A_l/B_l}_{\text{vz}}$ 
and $\lambda^{A_l/B_l}_{\text{ex}}$.
We denote by $s_{z}$ the spin Pauli matrix. Each $\hat{h}^l$ is a $4\times 4$ matrix in the spin-sublattice resolved basis $(A_{l\uparrow},A_{l\downarrow},B_{l\uparrow},B_{l\downarrow})$ within layer $l$ (bottom is $l=1$,
top $l=2$). Since spin-orbit coupling in graphene induced
by TMDCs is of the valley-Zeeman type \cite{Gmitra2015}, we set $\lambda^{A_l}_{\text{vz}} \approx 
-\lambda^{B_l}_{\text{vz}}$. 

The numerical values for the parameters of the single-particle Hamiltonian, 
$\hat{h}(\textbf{k},\tau)=\hat{h}_0(\textbf{k},\tau) + \hat{h}_{\text{prox}}(\tau)$,
are taken from the \emph{ab initio} results of Ref.~\cite{PhysRevLett.125.196402} and presented in Table~\ref{tab:bbg-prox}.
The calculated
low-energy band dispersions for WS$_2$/BBG/CGT are shown in Fig.~\ref{fig:scheme}(c).  
The signs of $V$ and electron doping $n_e$ determine the dominant proximity 
spin interaction. If both $V$ and $n_e$ positive or negative, the electrons at the Fermi level
experience proximity exchange coupling, while if the signs of $V$ and $n_e$ are opposite, 
the electrons at the Fermi level have strong proximity spin-orbit coupling; we use labels (SO) 
and (EX) in Fig.~\ref{fig:scheme}(c) and below when needed to keep track of the dominant spin interaction.

To investigate correlation phenomena we introduce the many-particle Hamiltonian
operator as the sum of the kinetic and potential energies 
$\hat{H}=\hat{H}_{\text{kin}}+\hat{H}_{\text{int}}$, where
\begin{align}
    \hat{H}_{\text{kin}}&=\sum_{\textbf{k}\tau s,ij}
    \hat{c}^{\dagger}_{s\tau i}(\textbf{k})
    \left[ \hat{h}(\mathbf{k},\tau) - \mu\right]_{si,s'j}\hat{c}_{s'\tau j}(\textbf{k}),
    \label{eq:kin}\\
    \hat{H}_{\text{int}}&=U_0\left(\hat{n}_{\uparrow K}\,\hat{n}_{\downarrow K}+\hat{n}_{\uparrow K^{\prime}}\,\hat{n}_{\downarrow K^{\prime}}\right)+
    U_1 \hat{n}_{K}\,\hat{n}_{K^{\prime}}.
    \label{eq:int}
\end{align} 
Here, $\hat{c}^{(\dagger)}_{s\tau i}(\textbf{k})$ 
is the annihilation (creation) operator for a Bloch electron with spin $s=\uparrow\hspace{-1.22mm}/\hspace{-1.22mm} \downarrow$ in valley $\tau={K/K^\prime}$ on BBG sublattice $i$ with valley-momentum $\textbf{k}$; $\mu$ is the chemical potential. 
The intra- and intervalley density interactions are described by repulsive (positive) couplings $U_0$ and $U_1$, respectively, while $\hat{n}_{s\tau}=\sum_{|\textbf{k}|<k_{c}}\sum_{i}\hat{c}^{\dagger}_{s\tau i}(\textbf{k})\hat{c}^{\phantom{\dagger}}_{s\tau i}(\textbf{k})$ stands for the spin-valley number operator in the $s\tau$-channel cut-off by a momentum $k_{c}$; the valley number operator then is 
$\hat{n}_{\tau}=\hat{n}_{\uparrow\tau}+\hat{n}_{\downarrow\tau}$. 

In the following, we consider SU(4)-symmetric interactions by setting 
$U_0 = U_1 = U=19$~eV~\cite{You2019, PhysRevB.105.134524, zhumagulov2023emergent}, and $k_{c}=0.06~\AA^{-1}$, which are consistent with the experimental pristine BBG phase diagram~\cite{Zhou2022, Zhang2023}. We use the same interaction parameters for the WS$_2$/BBG/CGT heterostructure. 

\section{Methodology.} 

\begin{table*}[t]
  \centering
\begin{tabular}{|c|cc|cc|cc|}
\hline 
pristine BBG phases & 
\multicolumn{2}{c|}{IVC}&
\multicolumn{2}{c|}{Stoner}\\
\hline
proximity phases & 
\multicolumn{1}{c|}{SVC$_{\pm}$} &
\multicolumn{1}{c|}{VC$_{\pm}$} &
\multicolumn{1}{c|}{SVP$_{\pm}$} &
\multicolumn{1}{c|}{VP$_{\pm}$}  \\
\hline 
spin-valley matrices $M_\Phi$  & 
\multicolumn{1}{c|}{
\makecell{$\left[s_x\tau_x\pm s_y\tau_y\right]$}
}  &  
\multicolumn{1}{c|}{
\makecell{$\left[s_0 \pm s_z\right]\tau_x$}
}     & 
\multicolumn{1}{c|}{
\makecell{$\left[s_z\tau_0\pm s_0\tau_z\right]$}
} & 
\multicolumn{1}{c|}{
\makecell{$\left[s_0 \pm s_z\right]\tau_z$}
}\\
\hline
\end{tabular}
\caption{
List of phases and corresponding phase operators $\hat{\Phi}=\sum_{|\mathbf{k}|<k_c}\sum_{i} \hat{c}^\dagger_{s\tau i}(\mathbf{k})\,\left[\,M_\Phi\,\right]_{s\tau,s'\tau'}\,\hat{c}^{\phantom{\dagger}}_{s'\tau' i}(\mathbf{k})=\hat{\Phi}^\dagger$ resolved in different spin-valley channels for the relevant symmetry-broken phases of pristine and proximitized BBG. The spin-valley resolved matrices $M_\Phi$ entering $\hat{\Phi}$'s are given in the third line. Phases involving $\tau_x$ or $\tau_y$ are valley mixed, i.e., they possess spatial modulation with vector $\mathbf{q}=(2\pi/3a)(1,\sqrt{3})$ which connects $K$ and $K^\prime$ valleys.
}
\label{tab}
\end{table*}

To resolve the correlated phases of our models of pristine and proximitized BBG we employ the RPA. First, we evaluate the non-interacting generalized static Lindhard's susceptibility, $\chi^{0}$,~\cite{PhysRevLett.101.087004, Graser2009, PhysRevB.83.100515} using the BBG Hamiltonian $\hat{H}_{\text{kin}}$, Eq.~(\ref{eq:kin}), at 0.4~K, considering 
different spin-valley modes $s\tau$ in the particle-hole channel. 
Second, we calculate the dressed susceptibility, $\chi=\left[1-\chi^{0}\Gamma\right]^{-1}\chi^{0}$, where $\Gamma$ is the irreducible vertex function corresponding to $\hat{H}_{\text{int}}$, Eq.~(\ref{eq:int}). 
Finally, we check for the divergence of $\chi$: If the highest eigenvalue $\lambda_c$ of $\chi^{0}\Gamma$ becomes greater or equal to unity, a correlated phase corresponding to that eigenvalue emerges.  

In particular, we find all possible correlated phases $\hat{\Phi}$ of different spin-valley channels by first diagonalizing $\chi^{0}\Gamma$ at $\mu = 0$ and $V = 0$. 
Then, varying the doping and displacement field, we compute the corresponding $\chi^{0}$ and $\chi^{0}\Gamma$. For each $\hat{\Phi}$ found above we estimate the critical parameter $\lambda_c(\hat{\Phi})=\langle\hat{\Phi}|\chi^{0}\Gamma|\hat{\Phi}\rangle/{\lVert \hat{\Phi} \rVert^2}$. The dominant instability is realized by phase $\hat{\Phi}$ that at given
$\mu$ and $V$ has the highest $\lambda_c(\hat{\Phi})$. 
Each symmetry-breaking phase $\hat{\Phi}$ can be expressed in terms of $\hat{c}^{(\dagger)}_{s\tau i}(\textbf{k})$ operators,
\begin{equation}
\hat{\Phi}=\sum_{|\mathbf{k}|<k_c}\sum_{i} \hat{c}^\dagger_{s\tau i}(\mathbf{k})\,\left[\,M_\Phi\,\right]_{s\tau,s'\tau'}\,\hat{c}^{\phantom{\dagger}}_{s'\tau' i}(\mathbf{k}),   
\end{equation}
where the 
spin-valley resolved matrices $M_\Phi$ that are relevant for our model are listed in Table~\ref{tab}.
However, at the phase boundaries, 
the dominant phase can be unresolved due to the degeneracy of several $\lambda_c(\hat{\Phi})$.  

\begin{figure}
    \centering
    \includegraphics[width=.5\textwidth]{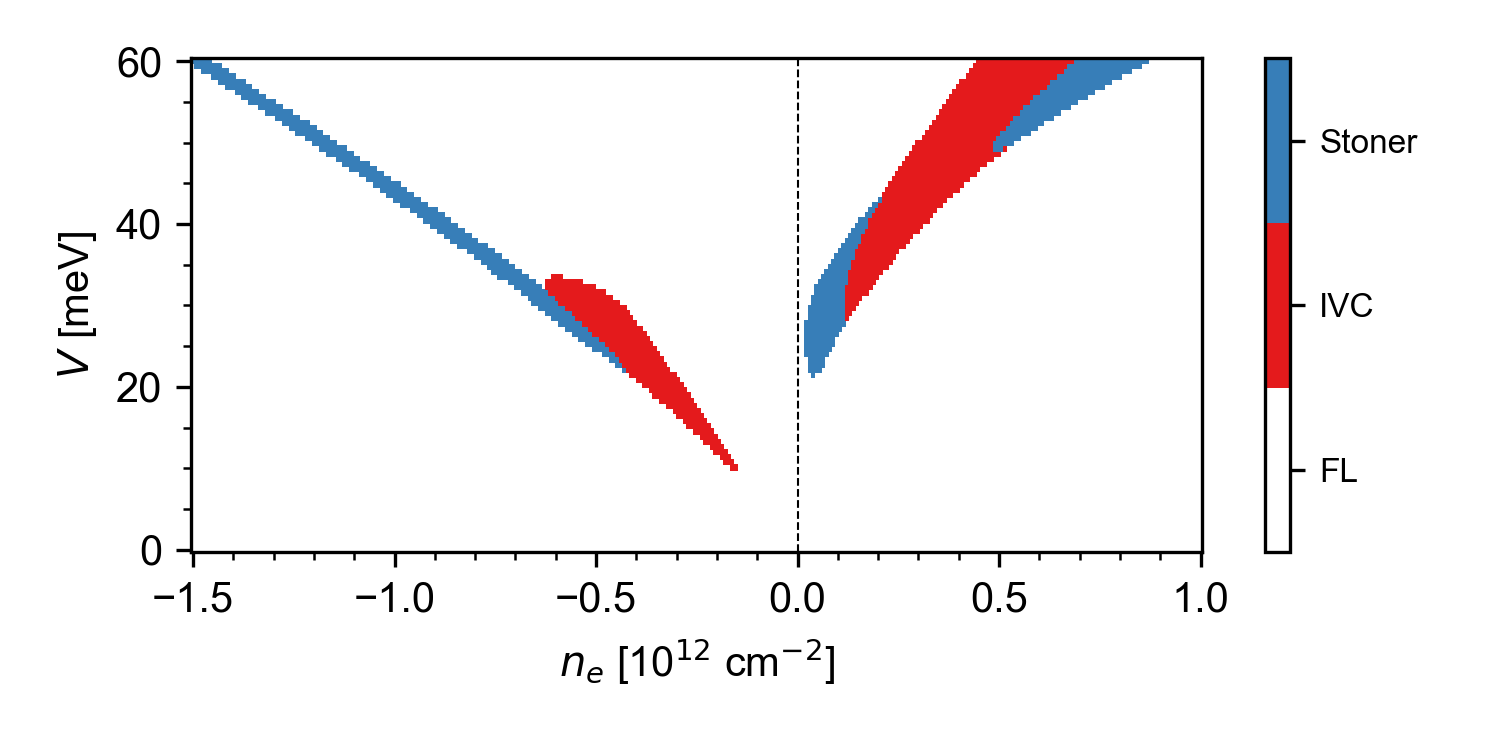}
    \caption{Calculated phase diagram of pristine BBG for varying doping $n_e$ and displacement field $V$. 
    We only show phases for $V>0$, as the diagram has $V$ to $-V$ symmetry.
    There are two dominant phases: the intervalley coherent phase (IVC), displayed by blue, and 
    the Stoner instability, shown in red. The white background corresponds to a stable Fermi liquid (FL). The phase diagram was obtained for $T=0.4$~K.}
    \label{fig:free}
\end{figure}

\begin{figure}
    \centering
    \includegraphics[width=.5\textwidth]{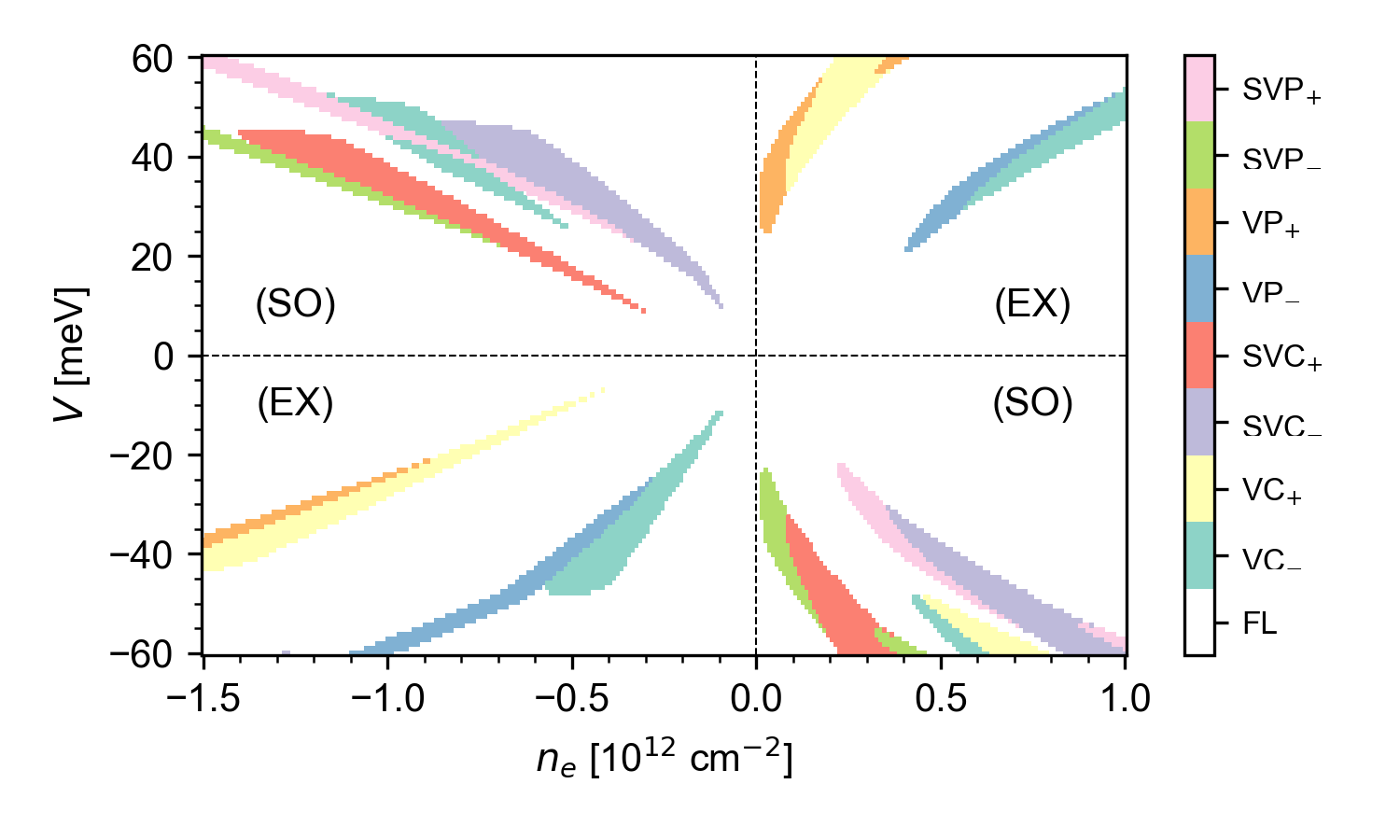}
    \caption{
    Correlated-phase diagram of a model ex-so-tic WS$_2$/BBG/CGT heterostructure, for 
    varying doping $n_e$ and displacement field $V$. 
    Eight symmetry-breaking phases are predicted to emerge due to the interplay of the
    electron-electron interactions and proximity-induced SO and EX splittings: VC$_{\pm}$ (valley coherence), SVC$_{\pm}$ (spin-valley coherence), SVP$_{\pm}$ (spin-valley polarized state), and VP$_{\pm}$ (valley polarized state), each marked by a different color. 
    The stable Fermi liquid phase is displayed by white. The phase diagram was obtained for $T=0.4$~K.
    }
    \label{fig:exc}
\end{figure}


Figure~\ref{fig:free} displays the obtained correlated phase diagram of pristine BBG. We
find IVC and Stoner instabilities, as also reported earlier~\cite{PhysRevB.105.134524, Chatterjee2022}. While the Stoner instability is local, the IVC state corresponds to a phase spatially
modulated by wave vector $\mathbf{q}$ which connects $K$ and $K^\prime$ valleys. 
The fact that $\hat{H}_{\text{kin}}$ lacks particle-hole symmetry is reflected in the
different phases for the electron and hole doping ranges. Both IVC and Stoner phases exhibit degeneracies, as listed in Table I, due to not fully broken spin-valley symmetries. 
The critical temperatures for each correlated phase depend heavily on the doping parameters and displacement field. With the chosen interaction parameters, it can reach up to 2.2~K for IVC and 1.0~K for the Stoner phases, as shown in Appendix ~\ref{app:temp}.

Figure~\ref{fig:exc} shows the main result of the paper. There, we plot the correlated phase diagram of 
WS$_2$/BBG/CGT for different doping levels and displacement fields. Compared with the pristine 
case, the phase diagram of proximitized BBG is rather rich.  
Indeed, if the Fermi level crosses bands experiencing strong proximity SO coupling, 
the Stoner phase breaks up into two SVP$_{\pm}$ 
states---both displaying spin-valley polarization along the $\pm z$ direction---while IVC splits into 
spin-valley coherent states SVC$_{\pm}$. 
Resolving specific spin-valley states within the Stoner and IVC phases is attributed to the
valley-Zeeman SO coupling~\cite{Gmitra2015}. 
Similarly, if the Fermi level lies in bands with strong EX coupling, 
the Stoner phase evolves into valley-polarized states, VP$_{\pm}$, while IVC 
is split into charge density waves, VC$_{\pm}$. The multitude of spin-valley-split 
phases of proximitized BBG due to SO and EX interactions are summarized in Table~\ref{tab},
including the explicit forms of the phase operators $\hat{\Phi}$. 
Their expectation values in the corresponding ground states can serve as the conventional order parameters discriminating different phases. 

Let us take a closer look at vHS. If the singularity is split due to the valley-Zeeman
SO coupling, two SVC$_{\pm}$ phases 
appear. As the corresponding $M_{\text{SVC}_{\pm}}$ contain spin $s_x$ and $s_y$ matrices, 
and valley $\tau_x$ and $\tau_y$ matrices,
they can be described as inter-valley spin-flip hopping that retains the spin-valley quantum number,
i.e., the product of $s \tau$. Say, spin up at $K$ is degenerate with spin down at $K'$. Coupling the two enabled by 
the Coulomb interaction lowers the kinetic energy, resulting in an SVC correlated phase. 
Similarly, when vHS are split by EX interaction, two spin-polarized VC$_{\pm}$ phases form up. Contrary to SVC$_{\pm}$, $M_{\text{VC}_{\pm}}$ does not mixes spins---it involves $s_0$ and $s_z$ spin matrices---but similarly as SVC$_{\pm}$ it intertwines the valleys---matrix $\tau_x$. 
Because of that both SVC$_{\pm}$ and VC$_{\pm}$ possess spatial modulations; the correlations 
effectively enlarge the BBG unit cell into a ``magnetic'' $3\times3$ - unit cell. Consequently, the Brillouin zone gets smaller while both valleys fold into the $\Gamma$ point. 
In what follows, the $\mathbf{k}$ vector is measured with respect to the center of such 
the reduced Brillouin zone. 

Finally, to find the effective single-particle excitation energies of the correlated phases
of our WS$_2$/BBG/CGT model, we employ the Hartree-Fock (HF) method and solve the following eigenproblem:
\begin{align}
\left(\hat{H}_{\text{kin}}+\hat{\Sigma}\right)|\Tilde{u}_n(\mathbf{k})\rangle=\Tilde{\varepsilon}_{n\textbf{k}}|\Tilde{u}_n(\mathbf{k})\rangle,
\end{align}
where $\Tilde{\varepsilon}_{n\textbf{k}}$ and $|\Tilde{u}_n(\mathbf{k})\rangle$ are HF-corrected quasi-particle energies and wave-functions labeled by the band index $n$ and the $\mathbf{k}$-vector in the reduced Brillouin zone. Correspondingly, $\hat{\Sigma}$ represents the HF self-energy, which mixes spin and valley indices; for details, see Appendix~\ref {app:hf}.

To illustrate how our ex-so-tic heterostructure enables to swap two correlated phases, one induced
by valley-Zeeman SO and the other by EX coupling, we consider
one particular doping level, namely hole density $n_e=-0.183\cdot10^{12}$~cm$^{-2}$. At positive $V= 18.5$~meV, the most stable correlated phase (largest $\lambda_c$) is a spin-valley coherence SVC$_{-}$. This phase arises due to the proximity spin-orbit coupling from WS$_2$, and couples the opposite spins at the two valleys. 
It is characterized by the flat band states close to the Fermi level exhibiting reduced spin polarization, which is not exactly zero because of the residual proximity exchange. 
Flipping the direction of the displacement field 
to $V=- 18.5$~meV, a different correlated phase emerges, VC$_-$, in which the states at the Fermi level 
are fully spin polarized, due to the proximity exchange from CGT.

\begin{figure}[h!]
    \centering
    \includegraphics[width=.5\textwidth]{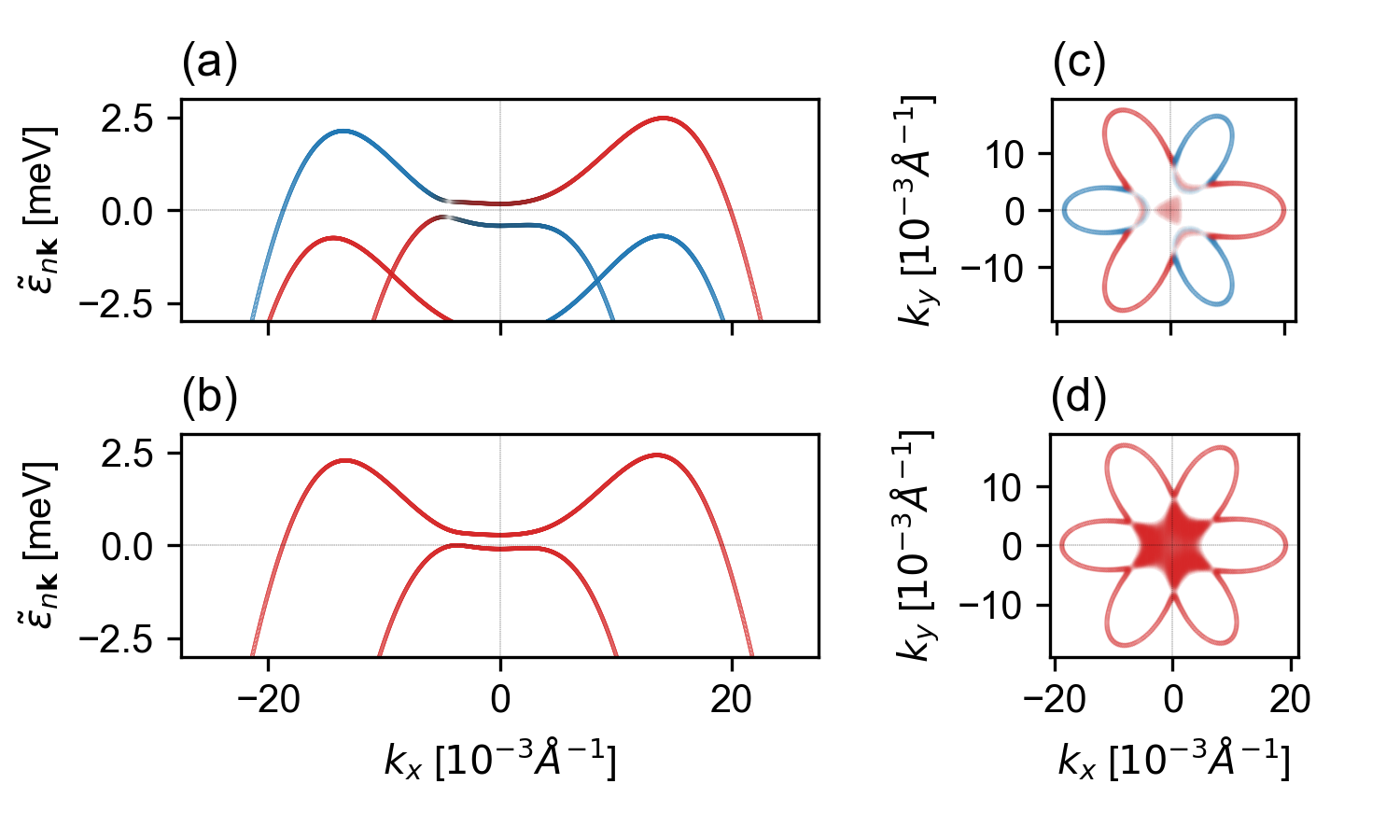}
    \caption{
    Swapping SVC and VC. Calculated HF quasi-particle band structures and spin-valley-resolved Fermi surfaces for the correlated SVC$_{-}$ phase, panels (a)~and~(c), with $V=+18.5$~meV and $n_e=-0.183\cdot10^{12}$~cm$^{-2}$ and VC$_{-}$ phase, panels~(b)~and~(d), with~$V=-18.5$~meV and $n_e=-0.183\cdot10^{12}$~cm$^{-2}$. 
    Red, blue, and grey lines mark, correspondingly, spin-up, spin-down, and spin-unpolarized states. 
    }
    \label{fig:fermi}
\end{figure}

In Fig. \ref{fig:fermi} we also show the markedly different quasi-particle Fermi contours of the two swappable phases SVC$_{-}$ and VC$_{-}$. To make this visualization, 
we plotted the thermally broadened density of states (derivative of the Fermi-Dirac function) at $T=0.4$\,K:
\begin{equation}
    \rho_{n\textbf{k}}=-\frac{d f(\varepsilon)}{d\varepsilon}\Bigl|_{\Tilde{\varepsilon}_{n\textbf{k}}}=\frac{1}{k_BT} \left[4\cosh\left(\frac{\Tilde{\varepsilon}_{n\textbf{k}}- \mu}{2k_BT}\right)\right]^{-2}.
\end{equation}

Inspecting Fig.~\ref{fig:fermi} we see that the quasi-particle band structures for the two considered phases, and also the shapes of their Fermi contours demonstrate very pronounced spectral asymmetries 
at the corresponding Fermi levels. 
However, regions with more smeared portions of the Fermi surface (different from just contour plots) are 
visible in the centre of the reduced Brillouin zone.

Knowing the HF self-energies $\hat{\Sigma}$ we are able to estimate the magnitudes of the associated correlated gaps, 
$\Delta_{\hat{\Phi}}=\sum_{s\tau,s'\tau'}\hat{\Phi}^{*}_{s'\tau',s\tau}\hat{\Sigma}_{s\tau,s'\tau'}/{\lVert \hat{\Phi} \rVert^2}$. This yields for the considered SVP$_{-}$ and VC$_{-}$ phases the following values: 
$\Delta_{\text{SVP}_{-}}=0.109$~meV and $\Delta_{\text{VC}_{-}}=0.085$~meV. 

To model an interacting system with a symmetry-broken phase at a self-consistent mean-field level, one can factorize $\hat{H}_{\text{int}}$, Eq.~(\ref{eq:int}), in terms of accessible $\hat{\Phi}$'s and for a given dominant phase approximate the interaction just by $\hat{H}_{\Delta}=\Delta_{\hat{\Phi}}\hat{\Phi}$.
For example, for a SVC$_{\pm}$ phase, one explores an effective Hamiltonian $\hat{H}_{\text{kin}}+\Delta_{\text{SVC}_{\pm}}\hat{\Phi}_{\text{SVC}_{\pm}}$
which promotes spin-flip-valley-flip hopping due to the interplay of intervalley electron-electron interactions and the valley-Zeeman SO coupling. Such a phase enables novel spin interactions and
is expected to result in new phenomena in spintronics \cite{RevModPhys.76.323}. 
In turn, a VC$_{\pm}$ phase can be treated on the mean-field level by
$\hat{H}_{\text{kin}}+\Delta_{\text{VC}_{\pm}}\hat{\Phi}_{\text{VC}_{\pm}}$
which gives rise to spatial modulations of the already spin-split bands due to the proximity-induced 
EX interaction. 

\section{Conclusion.} By performing realistic calculations at the RPA level, and by computing Hartree-Fock excitation spectra, we predict that the correlated states of ex-so-tic heterostructures
based on BBG can be swapped between SO and EX-driven phases. A single device can exhibit the full spectrum
of correlated phases, from uniform Stoner valley polarized, to spatially modulated spin-polarized VC and 
spin-flip-valley-flip spin-valley coherences. 
While we specifically consider the DFT parameterization of one stacking of WS$_2$/BBG/CGT, 
our findings are more general and valid for twisted structures with modified proximity spin interactions, as well as for different encapsulating spin-orbit materials and magnetic semiconductors/insulators. 

\begin{acknowledgments}
This work was funded by the Deutsche Forschungsgemeinschaft (DFG, German Research Foundation) 
SPP 2244 (Project No. 443416183), SFB 1277 (Project-ID 314695032), by the European Union Horizon 2020 Research and Innovation Program under contract number 881603 (Graphene Flagship), and by FLAG-ERA project 2DSOTECH.
D.K.~acknowledges partial support from the IM\-PULZ project IM-2021-26---SUPERSPIN funded by the Slovak Academy of Sciences and VEGA-grant DESCOM 2/0183/21.
\end{acknowledgments}

\appendix

\section{Generalized Lindhard's susceptibility tensor.}
\label{app:lind}
To calculate the static susceptibility tensor, $\chi^{0}_{abcd}$, we use the generalized Lindhard's susceptibility formula that employs the eigenstates and eigenvalues of the full single-particle 
Hamiltonian $\hat{h}_{\tau s}(\textbf{k})$, in the case of pristine or proximitized BBG, correspondingly, with or without the proximity-furnished spin-orbit and exchange interactions:
\begin{align}
    \sum_{j}\hat{h}^{a}_{ij}(\textbf{k})u^{aj}_{n\textbf{k}}=\varepsilon^{a}_{n\textbf{k}}u^{ai}_{n\textbf{k}},
\end{align}
where $\varepsilon^{a}_{n\textbf{k}}$ denotes single-particle energy and $u_{n\textbf{k}}^{ai}$ 
is amplitude of the $n$-th eigenstate $|u_{n\textbf{k}}\rangle$ projected onto a sublattice $i$ in a spin-valley channel $a$. 
In what follows we reserve indices $m,n$ to label electronic bands stemming from the single-particle Hamiltonian $\hat{h}(\textbf{k})$, and indices $i,j$ to label BBG sublattices $(A_1,B_1,A_2,B_2)$. 
Moreover, we introduce four quantum numbers $0,1,2,3$---overrun by indices $a,b,c,d,u,v$---that 
overall parameterize the spin-valley degrees of freedom:
\begin{align}
    &0=\left(\uparrow\times K\right),
    1=\left(\uparrow\times K^{\prime}\right),\\
    &2=\left(\downarrow\times K\right),
    3=\left(\downarrow\times K^{\prime}\right).\nonumber
\end{align}
Also, we would like to point out that indexes with a bar $\bar{a},\bar{b}$, etc. correspond to creation operators, while those without a bar $a,b$, etc. correspond to annihilation operators.

The generalized Lindhard's susceptibility tensor, $\chi^{0}$, can be given in terms of eigenenergies $\varepsilon^{a}_{n\textbf{k}}$ and projections $u_{n\textbf{k}}^{ai}$
as follows:
\begin{align}
    \chi^{0}_{\bar{a}b\bar{b}a}=&
    A_{uc}\int_{|\textbf{k}|<k_{c}}\frac{\mathrm{d}^2\textbf{k}}{(2\pi)^2}\, \\
    &\sum_{nm}\sum_{ij}
    \frac{f^{a}_{n\textbf{k}}-f^{b}_{m\textbf{k}}}{\varepsilon^{b}_{m\textbf{k}}-\varepsilon^{a}_{n\textbf{k}}}
    (u_{n\textbf{k}}^{ai})^{*}
    u_{m\textbf{k}}^{bi}
    (u_{m\textbf{k}}^{bj})^{*}
    u_{n\textbf{k}}^{aj}\,,\nonumber
\end{align}
where $A_{uc}$ stands for an area of the unit cell and the corresponding Fermi-Dirac weights
$f^{a}_{n\textbf{k}}$ read:
\begin{equation}
f^{a}_{n\textbf{k}}=\frac{1}{1+\exp[\left(\varepsilon^{a}_{n\textbf{k}}-\mu\right)/k_BT]}\,.
\end{equation}
where $\mu$ is chemical potential.

The static susceptibility, $\chi^{0}$, as used in the main text and below is computed for $T = 0.4$~K (equivalent in energy to $0.02$~meV what is order of magnitude below the magnitudes of all estimated correlated gaps $\Delta_{\hat{\Phi}}$), and for the momentum cutoff $k_{c}= 0.06~\AA^{-1}$. 
The latter corresponds to the orbital-energy scale $\sqrt{3}/2a\gamma_0 k_{c} = 0.453$~eV (in this expression, unluckily, $a$ stands for the graphene lattice constant and not for the spin-valley index), i.e.~the employed ultraviolet cutoff is sufficiently larger than the spin-orbit and exchange coupling strengths, displacement field magnitudes, and chemical potentials as considered in this work. This proves that the used temperature $T$ and electron-electron couplings $U_0$ and $U_1$ are chosen meaningfully 
and are not spoiling the physical results obtained in our calculations.

\section{Random Phase Approximation: Vertex Function.}
\label{app:rpa}

We treat emergent correlations by mean of the following density-density Hamiltonian:
\begin{align}    
\hat{H}_{\text{int}}&=
U_{0} (n_{\uparrow K} n_{\downarrow K} +n_{\uparrow K^{\prime}} n_{\downarrow K^{\prime}})
+
U_{1} n_{K}n_{K^{\prime}}\\
&=
U_{0}(n_0 n_2+n_1 n_3) + U_{1}(n_0+n_2)(n_1+n_3)
\nonumber
\label{eq:int}
\end{align}
where $U_{0}>0$ and $U_{1}>0$ are parameterizing the intra- and intervalley effectively repulsive interactions.
To obtain the RPA vertex function, $U$, we use the convention
, which gives:
\begin{equation}
    \hat{H}_{\text{int}}=\frac{1}{4}\sum_{abcd}U_{c\bar{a}d\bar{b}}
    \hat{c}^{\dagger}_{\bar{a}}
    \hat{c}^{\dagger}_{\bar{b}}
    \hat{c}_{c}
    \hat{c}_{d}
\end{equation}
The representative components of this tensor are given by,
\begin{align}
    U_{0220}&=U_{1331}=+U_{0}\\
    U_{0022}&=U_{1133}=-U_{0}\nonumber\\
    U_{0110}=U_{0330}&=U_{1221}=U_{2332}=+U_{1}\nonumber\\
    U_{0011}=U_{0033}&=U_{1122}=U_{2233}=-U_{1}\nonumber
\end{align}
All unmentioned components of the $U$ tensor can be obtained by the relation $U_{a\bar{b}c\bar{d}}=U_{d\bar{c}b\bar{a}}$.

\section{Self-consistent Hartree-Fock calculations.}
\label{app:hf}
This work employed a self-consistent Hartree-Fock approach to determine the correlated band structure. The effective Schrödinger equation is given by:
\begin{align}
\sum_{b}\left[\varepsilon^{a}_{n\textbf{k}}\delta_{\bar{a}b}+\Sigma_{\bar{a}b}\right]\Tilde{u}_{n\textbf{k}}^{b}=\Tilde{\varepsilon}_{n\textbf{k}}\Tilde{u}_{n\textbf{k}}^{a},
\end{align}
where $\Tilde{\varepsilon}^{a}_{n\textbf{k}}$ and $\tilde{u}_{n\textbf{k}}$ denote Hartree-Fock corrected single-particle energy and wavefunction, as before $i$ denote band index and $a,b$ is spin-valley degree of freedom. 
$\Sigma_{ab}$ is the Hartree-Fock self-energy that was calculated as follows:
\begin{align}
    \Sigma_{\bar{a}b}=-\sum_{cd}U_{b\bar{a}c\bar{d}}G_{c\bar{d}},
\end{align}
where $U_{abcd}$ is the RPA vertex function and $G_{cd}$ is the density matrix. The density matrix is subsequently determined as follows:
\begin{align}
    G_{c\bar{d}}=A_{uc}\int_{|\textbf{k}|<k_c}\frac{\mathrm{d}^2\textbf{k}}{(2\pi)^2}\,\sum_{i,n}\left[\Tilde{f}_{n\textbf{k}}-\frac{1}{2}\right]
    \Tilde{u}_{n\textbf{k}}^{ci}
    (\Tilde{u}_{n\textbf{k}}^{di})^{*}
    \label{eq:green}
\end{align}
Note that the orbital and band indices are traced out to calculate the density matrix $G_{cd}$. The corresponding Fermi-Dirac weights
$\Tilde{f}_{n\textbf{k}}$ read:
\begin{equation}
\Tilde{f}_{n\textbf{k}}=\frac{1}{1+\exp[\left(\Tilde{\varepsilon}_{n\textbf{k}}-\mu\right)/k_BT]}\,.
\end{equation}

We constrain chemical potential $\mu$ by fixing electron doping: 
\begin{align}
    n_e=\int_{|\textbf{k}|<k_c}\frac{\mathrm{d}^2\textbf{k}}{(2\pi)^2}\,\sum_{n}\left[\Tilde{f}_{n\textbf{k}}-\frac{1}{2}\right],
    \label{eq:den}
\end{align}
The offset $-1/2$ in equations (\ref{eq:green}) and (\ref{eq:den}) is intended to fix the chemical potential equal to zero at the point of charge neutrality.

\section{Two-particle response function linear algebra.}
\label{app:two}
In our work, we used the following linear algebra rule for the product of two-particle response functions in a particle-hole channel:
\begin{align}
    \left[A \cdot B\right]_{abcd} =\sum_{uv} A_{abuv}B_{vucd}
\end{align}
Thus, the expression for calculating the critical parameter can be obtained as follows:
\begin{align}
    \lambda_c=\frac{1}{\lVert \hat{\Phi} \rVert^2}\sum_{ab}\sum_{cd} \hat{\Phi}^{*}_{b\bar{a}}\chi^{0}_{\bar{a}b\bar{b}a}U_{a\bar{b}c\bar{d}}\hat{\Phi}_{\bar{d}c}
\end{align}

The most relevant set of $\hat{\Phi}$'s corresponding to correlated phases in BBG with and without spin-orbit and exchange couplings was obtained by the diagonalization of the symmetric matrix $\chi^{0}U$:
\begin{align}
    \lambda_c\hat{\Phi}_{\bar{a}b}=\sum_{cd}\chi^{0}_{\bar{a}b\bar{b}a}U_{a\bar{b}c\bar{d}}\hat{\Phi}_{\bar{d}c},
\end{align}
at charge neutrality point and zero displacement field. The list of $\hat{\Phi}$'s is provided in Table~1 in the main text.

The value of the correlated gap in our work was calculated using the following formula:
\begin{align}
    \Delta_{\hat{\Phi}}=\frac{1}{\lVert \hat{\Phi}\rVert^2} \sum_{ab} \hat{\Phi}^{*}_{b\bar{a}}\Sigma_{\bar{a}b}
\end{align}
Prefactor $1/\lVert \hat{\Phi} \rVert^2$ is due to the normalization of $\hat{\Phi}$, as represented in Table 1 in the main text. 

\section{Temperature/doping phase diagram for pristine BBG.}
\label{app:temp}

\begin{figure}[!h]
    \centering
    \includegraphics[width=.5\textwidth]{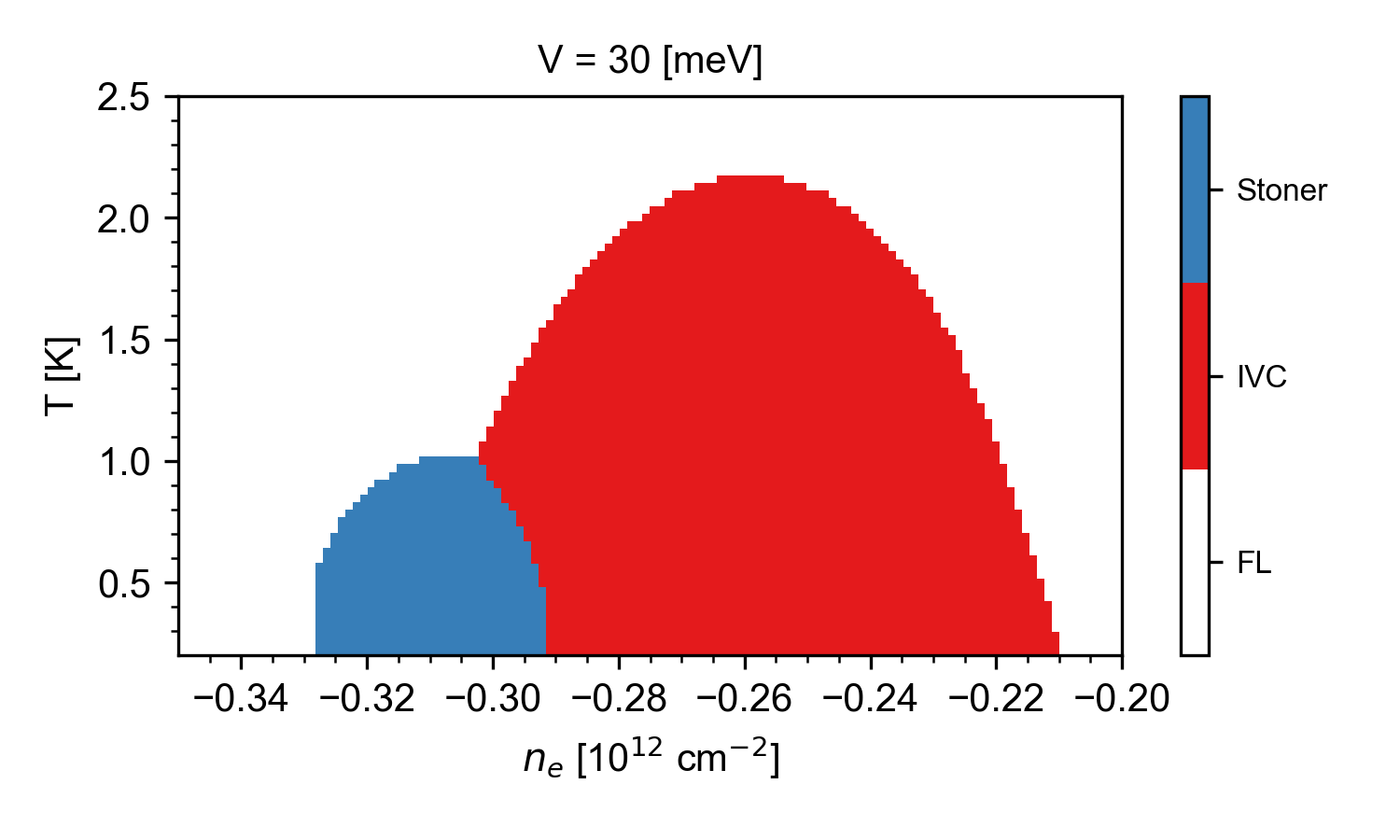}
    \caption{
    Temperature/doping phase diagram for pristine BBG at the displacement field with a layer energy difference of 30 meV.  There are two dominant phases: the intervalley coherent phase (IVC), displayed by blue, and the Stoner instability, shown in red. The white background corresponds to a stable Fermi liquid (FL). 
    }
    \label{fig:temp}
\end{figure}

To clarify the robustness of the different phases with temperature, we calculated a temperature/doping phase diagram for pristine BBG at the fixed displacement field $V=30$~meV, which is displayed in Fig.~\ref{fig:temp}. The Stoner phase is predicted to be stable up to 1.0~K, while IVC can be up to 2.2~K.

\end{document}